# A Heuristic Algorithm for optimizing Page Selection Instructions

Qing'an Li[1], Yanxiang He[1,2], Yong Chen[1], Wei Wu[1], Wenwen Xu[3]

1. Computer School, Wuhan University, Wuhan 430072, China
2. State Key Laboratory of Software Engineering, Wuhan University, Wuhan 430072, China
3. Institute of Computing Technology, Chinese Academy of Sciences

*Abstract*-**Page switching is a technique that increases the memory in microcontrollers without extending the address buses. This technique is widely used in the design of 8-bit MCUs. In this paper, we present an algorithm to reduce the overhead of page switching. To pursue small code size, we place the emphasis on the allocation of functions into suitable pages with a heuristic algorithm, thereby the cost-effective placement of page selection instructions. Our experimental results showed the optimization achieved a reduction in code size of 13.2 percent.**

## I. INTRODUCTION

In many kinds of RISC based MCUs, such as the PIC16F7X [1] family of Microchip Technology Inc, the No. 1 8-bit microcontroller manufacturer, the length of instructions is greatly limited. To support large programs, a special register is designed to store the high part of code's address. As a result, the program memory layout seems to be multi-paged, with each page of a fixed size. The page size is determined by the bits of an instruction used to store the code address. For example, if a function call instruction is 14 bits wide, among which 11 bits can be used to indicate the code address, the size of each page is 2K * 14 bits. So, to support program memory as large as 8k, this special register is required to provide at least 2 bits to indicate the high part of code address, sometimes called the page number. A multi-paged program memory with page size of 2K is illustrated in Figure 1. .

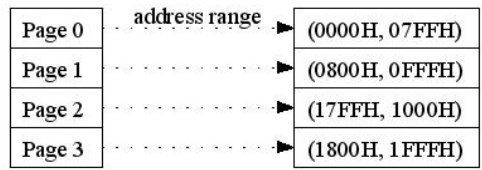

Figure 1.  Program Memory Maps

With this kind of multi-paged program memory layout, before the control flow is transferred from the present instruction to another far away, extra operations on this special register are necessary. The instructions related to these operations, called page selection instructions in this paper, inevitably induce extra overhead in code size. Code size is critical for the programs running in embedded systems, since smaller code size often means less consumption of ROM as well as less energy, and thus more competitiveness for IC manufacturers.

This paper presents an algorithm to optimize these page selection instructions by cost-effective allocation of functions and placement of page selection instructions. This paper is organized as follows. Our algorithm is discussed in detail in Section II. Our experimental results are shown in Section III. Related works are reviewed in Section IV. Then, a conclusion is drawn in Section V. Finally, Section VI lists the references for this paper.

## II. BACKGROUND

### A. Definitions

The following definitions are helpful for understanding this paper.

*1) Page*

Page is a logical concept rather than a physical one. As stated above, page size is determined by the bits in an instruction to indicate the code address. In PIC16F7X family, only 11 bits are used to indicate the address, so a page here is a space of sequential addresses, of which the start is dividable by 2K (2^11).

*2) Page Selection Register (PSR)*

In this paper, the special register used to indicate the page number, is called page selection register (which is named PCLATH in the PIC family).

*3) Page Selection Instruction (PSI):*

The instructions designed to switch the page number are called page selection instructions. That is, a PSI can write into the PSR directly.

*4) Value of PSR*

PSR is a register of 8-bit, but only two or three bits are used to indicate the page number. However, in this paper, the value of PSR is defined to be the page number indicated by PSR.

*5) Basic Block*

This is commonly used in the compiler optimization terminology [2]. Briefly, a basic block is a sequence of instructions in which flow of control can only enter from its beginning and leave at its end.

*6) Page Transparent Instruction (PTI)*

Generally, only the jump operation and function call operation trigger the loading of value of PSR into PC. The instructions related to these operations are called page nontransparent instructions (PNTI); others are page transparent. In this paper, only two kinds of instructions are page nontransparent.

*a) goto*

Before such an instruction executes, the value of PSR is required to be the same with the number of the page holding the current function.

*b) call*

Before a function call instruction executes, the value of PSR is required to be the same with the number of the page holding the callee function. After this instruction, PSR may be changed by PSIs in the callee function.

*7) Page Transparent Block (PTB)*

If a block includes any PSI, the block is called a page nontransparent block (PNTB); otherwise, the block is page transparent.

*8) FuncPage*

This function mapping from the functions to the page numbers, is used to indicate the number of page holding a certain function. For a function f, FuncPage (f) means the number of the page holding f.

### B. The Motivations

The motivation of our algorithm comes from the following observations. Firstly, the value of PSR at any point is always related to a function or more[1], since PSR indicates the page holding a related function. Besides, at any point immediately before a PNTI, a PSI is needed only when the current value of PSR is not the same with what this PTI requires. It follows that if the functions are located delicately, the chance that the current value of PSR is just what the PTI requires could be increased. An example is illustrated in Fig. 2. If the current value of PSR is FuncPage (f), and the value required is FuncPage (g), by placing f and g into the same page, no PSI is need. So the goal of the optimization is to minimize the PSIs to the most extent, and a feasible way to this goal is through placing the related functions in the same page as possible as we could. The more savings we could obtain by allocating two functions into one page, the more related they are viewed as. For example, the caller function is related with its first callee function, since allocating them into the same page could save the PSIs before the call instruction.

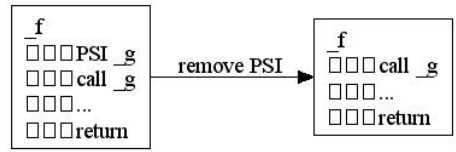

Figure 2.  Optimizing the PSI by function allocation

---

[1] There is a need to make clear that the value of PSR at some point may be related to more than one function, when more than one path can reach this point or the value is affected by a function call instruction.

## III. THE ALGORITHM

An algorithm is presented in this paper devoted to reduce the number of PSIs, with careful allocation of related functions into the same page. After this allocation, a cost-effective placement of PSIs is obtained. The placement of PSIs would not be described in detail, since once the function allocation is finished, the value of PSR required before each PNTI is determined, and what is left is to insert PSIs before a PNTI when the current value of PSR differs from the number of the required page. Therefore, our algorithm consists of only the following three steps.

The first step is to found the functions to which the value of PSR is related at every point of the analyzed code. For example, before a call instruction, the callee function is related to the value of PSR; after this call instruction, either the last function invoked (directly or indirectly) by the callee function, or the callee function itself, is related to the value of PSR. This step is called the analysis process. Then, we build a weighted function relation graph (FRG). In this FRG, once it is found that the placing of two functions into same page could result in savings of PSIs, the weight value of the edge between the two functions is increased. This step is called the building process. At last, given the number of pages, we place the functions of the analyzed code into the right pages by partitioning the weighted FRG. If we allocate the functions delicately, the number of PSIs can be reduced to the most extent. This step is called the partitioning process.

### A. The analysis process

In this process, we use the algorithm of data flow analysis [2] to calculate the set of functions related to the value of PSR at every point of the code. Firstly, we use the equations depicted in Figure 3. to calculate whether and how a block affects the value of PSR. Both Gen and Kill are sets of functions of the analyzed code. The function RetVop is depicted in Figure 5. It's obvious that only a PNTB has the potential to affects the value of PSR.

$Gen(b) = RetVop(i)$, if $b$ is a PNTB and $i$ is the last PNTI of $b$
$Gen(b) = \{\ \}$, if $b$ is a PTB
$Kill(b) = \{f \mid f \notin RetVop(i)\}$, if $b$ is a PNTB and $f$ is a function
$Kill(b) = \{\ \}$, if $b$ is a PTB

Figure 3.  Equations for calculating Gen and Kill sets

Then, we use the equations in Figure 4. to iteratively determine which block or blocks affect the value of PSR at the entry and exit point of each block. As Gen and Kill, both In and Out are sets of functions. Before the iteration progress begins, the In and Out sets for any block are initialized to be empty.

$$In(b_i) = \bigcup_{b_j \in Preds(b_i)} Out(b_j)$$
$$Out(b_i) = In(b_i) - Kill(b_i)$$

Figure 4.  Equations for calculating In and Out sets

Now, it is easy to calculate the set of functions related to the value of PSR at any point of the program by scanning each basic block only once. For a point immediately after instruction $i$, marked as $p_i$, the algorithm depicted in Fig. 5 can calculate the value of PSR at this point, marked as VOP (i). In this figure, Out (f.psb) means the Out set for the pseudo block of function f. A pseudo block is not a real block of code. After we construct the control flow graph (CFG), we add a pseudo block into the CFG such that it is connected to any block with no successor in the CFG, and it becomes their common successor. Therefore, this pseudo block is the unique exit of this function.

```
1    func GetVop
2       if pi is the entry point of a block b
3           VOP (pi) = In (b)
4       else if i is a PTI
5           VOP (i) = VOP (i-1)
6       else
7           VOP (i) = RetVop i

8    func RetVop i
9       if i is a "goto" instruction
10          return the current function
11      else if i is a "call f" instruction
12          return Out (f.psb)
```

Figure 5.    Calculating value of PSR at any point

It's noteworthy that there seems to be a cyclic dependency in this algorithm, since RetVop depends on Out, Out depends on Gen, and Gen depends on RetVop. There are two explanations. In the first place, even though there are cyclic function callings in the program, there is a fix point for the inter-procedural flow analysis [2]. In the second place, if there are no cyclic function callings in the program, a more efficient algorithm could be obtained by processing the callee function before the caller function. A topological sorting could do it well.

### B. The building process

With the analysis from the first step, it is easy to build the weighted function relation graph. Firstly, we build a complete graph, with each node representing a function and initialize the weight value of all the edges to be zero. Then, by scanning all the PNTIs in the program once, the weight value is updated with the algorithm depicted in Figure 6. After this step, it is assumed that the weight value of an edge between two functions could be approximately equated with the cost savings we could gain by placing these two functions in the same page.

### C. The partitioning process

Until now, all tasks left are about how to partition the FRG. With the assumption stated in the building process, the sum of the weight value of all edges in the FRG is the cost savings in total. Therefore, the problem of page selection optimization can be restated as follows:

 1) Any function could be placed into only one page.
 2) The number of pages to be used are speicfied by the MCU.
 3) The sum of size of functions placed in one page should never be greater than the size of the page.
 4) If two functions are placed into the same page, the weight value of the edge between the two functions can be reset to zero.
 5) The goal is to find such a partition that the four conditions above are fully satisfied, and the sum of the weight value of all edges in the reserved graph is minimized.

The partitioning problem has been proved to be NP hard, so there is no optimal polynomial algorithm. A greedy algorithm is presented in this paper, described in Figure 7. In this algorithm, the allocation is conducted by some dynamically updatable statistics. We always try to allocate a function to the right page, when this allocation could save the most PSIs from the statistics (decrease the most weight value from the current graph).

### D. Complexity analysis of the algorithm

For the first step, the Gen and Kill sets could be calculated by scanning the whole program once. Although iteration is needed in calculating the In and Out set, the times of iteration is bounded by a small constant factor [2]. So, time cost is linear to the code size $S$.

For the second step, most instructions are PNTIs and the VOP (i) includes all the functions in the worst case. If the number of

```
1    func AddWeight i, f
2    // Graph is the complete graph about function relation
3    // i is the current analyzed instruction, f is the current analyzed function
4    // PreValue is a predefined value for calculating the weight value
5    if i is a "goto" instruction
6        if VOP(i-1) is not equal to {f}
7            forall pair of functions (g,h) in {VOP (i-1)} U {f}
8                add PreValue/|VOP (i-1)| to Graph (g, h)
9    else if i is a "call e" instruction
10       if VOP(i-1) is not equal to {e}
11           forall pair of functions (g,h) in {VOP(i-1)} U {e}
12               add PreValue/|VOP (i-1)| to Graph ( g, h )
```

Figure 6.    Code for calculating weight value

```
1    func partition
2    // Pages is a list of pages, the number is specified by the MCU
3    // Funcs is a list of all the functions; Graph is the RFG
4    // Weight is a map from functions and pages to weight value, indicating the cost savings
5    initialize the size of each of the pages to 2k
6    sort the Funcs by size of function size in descend order
7    get the first function f from the Funcs
9    get a page p from the Pages
10   if p.size > f.size
11       remove f from Funcs
12       p.size <- p.size – f.size
13       forall g from Funcs
14           Weight[p,g] <- Weight[p,g] + Graph[f,g]
15   else   print "not enough memory error"
17   while Funcs are not empty
18       get the function f from the Funcs with Weight[p,f] is the greatest in Weight
19       calculate the number n of PNTIs in f and estimate the size f.size of f
20       if p.size > f.size
21           remove f from Funcs
22           p.size <- p.size – f.size
23           p.size <- p.size + n             // complementing the cost savings
24           forall g from Funcs
25               Weight[p,g] <- Weight[p,g] + Graph[f,g]
26       else if find page p' from Pages, with p'.size > f.size
27           p <- p'
28       else   print "not enough memory error"
```

Figure 7.    Code for function partitioning

functions is *NOF*, then the code in the 8th line and the 12th line in Figure 6. executes at most $NOF^2$ times. Therefore, the time cost is linear to $S*NOF^2$.

For the third step, the time cost is dominated by the code in the 25th line in Figure 7. If the number of the pages is *NOP*, and Weight and Graph are implemented with a random access data structure, the time cost is linear to $NOP*NOF^2$.

Since S is commonly greater than *NOP*, the time cost for this algorithm is linear to $S*NOF^2$. The space cost is dominated by the implementation of Weight and Graph, which respectively are bounded by *NOP*NOF* and *NOF*NOF*

## IV. EXPERIMENTAL RESULTS

We implemented the algorithm stated above in a cross compiler framework, HICC. HICC compiles source code written in C language into target code executable on the HR6P family of microcontrollers, among which each kind of microcontroller typically constitutes a RISC-based Harvard architecture with instruction sizes of 14, 15, or 16 bits, and a data-bus that is 8-bit wide. With a special register for the higher part of code size, also called PSR, these MCUs could support program memory of at most 64K*16 bits.

Our experiments were conducted on the HR6P90H MCU [4], which provides 8K or 16K * 15 bits of program memory. The benchmark suit comprises 21 programs from industrial applications in embedded systems, such as those for electric stoves, electric bicycles and washing machines. The result illustrated in Figure 8. showed that the code size shrank to 86.8 percent on average.

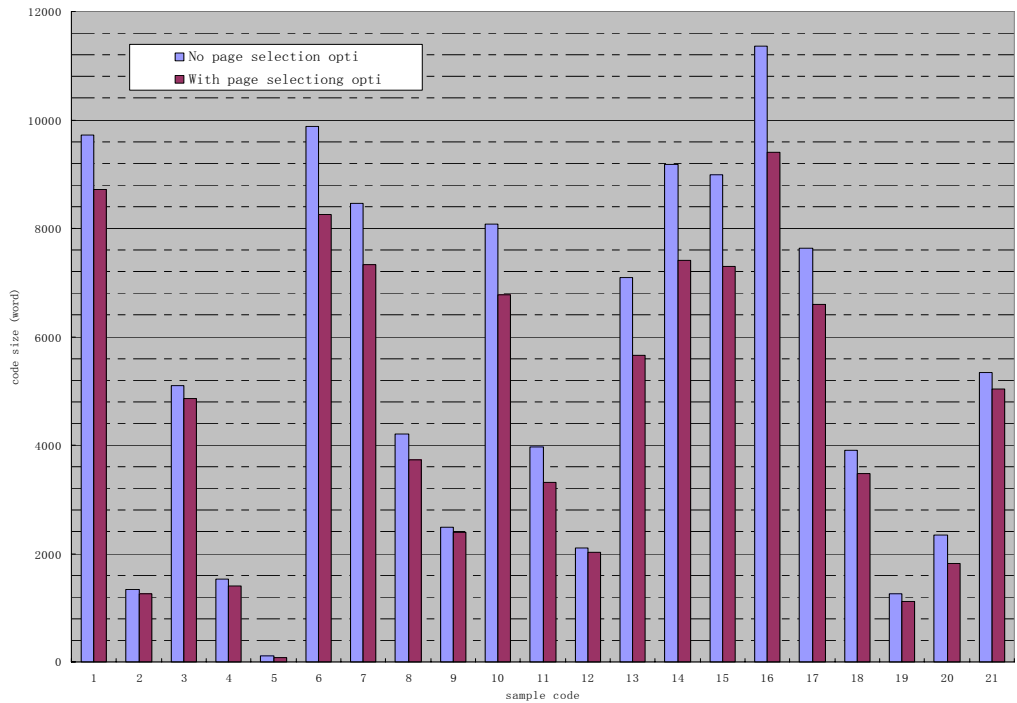

Figure 8. Code compiled with optimization and without optimization

## V. RELATED WORKS

To us know, there is little research in the literature about optimizing the page selection instructions, although the problem exists for a long time. However, many works have been done for optimizing bank selection instructions, which is very close to the problem of optimizing page selection instructions. Bernhard et al. [5] formulated the problem of optimizing bank selection instructions for multiple goals as a form of Partitioned Boolean Quadratic Programming (PBQP). However, they made the assumption that the variable had been assigned to specified banks. In the work by Yuan Mengting et al. [6], the variable partitioning was taken on a part of the memory, namely the shared memory, which is highly architecture dependent. Liu Tiantian et al. [7] claimed they had integrated variable partitioning into optimizing bank selection instructions. But, with the analysis of code patterns, they placed the emphasis on the positions for inserting bank selection instructions rather than the variable partitioning. Many other works [8] about variable partitioning focused on DSP processors, where parallelism and energy consumption attracted the main attentions. There is also research work to improve the overall throughput for MPSoc architecture by variable partitioning.

There are some differences between optimizing bank selection and page selection. The construction of the function relation graph is not as simple as the construction of the variable access graph, since the former may generally need inter-procedural analysis. Besides, the placement of a function could dynamically change size of the function itself, since some PSIs are optimized. This makes it more difficult to estimate the reserved space of a page.

VI. CONCLUSIONS AND FUTURE WORKS

In this paper, we present an algorithm to optimize the page selection instructions by function partitioning. Our experimental showed it achieved a great improvement with respect to code size. However, there are still much work to be done to improve this algorithm. Perhaps the followings are worthy of consideration.

- Analyze the code patterns in more detail, as inspired by [7]. Our assumption mentioned above is made roughly. The weight value should be estimated more precisely to conduct the partitioning process. Maybe a probability algorithm could work well in estimating weight value.

- The algorithm presented in this paper for partitioning process may be somewhat naive. We could try to design a more complex but efficient one. Perhaps a cluster algorithm is worth a try.